\documentclass[a4paper,10pt]{article}
\usepackage[pdftex]{graphicx}
\PassOptionsToPackage{hyphens}{url}\usepackage{hyperref}
\usepackage{amsmath,amsfonts,amssymb}
\usepackage{fullpage}
\usepackage{authblk}
\usepackage{setspace}
\usepackage{caption}
\usepackage{subcaption}
\usepackage{comment}
\usepackage{booktabs}
\usepackage{tabularx}
\usepackage{adjustbox}

\usepackage[explicit]{titlesec}
\usepackage{sidecap}
\usepackage{pbox}
\usepackage[superscript]{cite}

\usepackage{enumitem}
\usepackage{multirow}
\usepackage{longtable}
\usepackage{makecell}
\usepackage{xcolor}
\date{}
\title{Collective intelligence and the blockchain: \\ Technology, communities and social experiments}

\author[1,2,3,*]{Andrea Baronchelli}

\affil[1]{\small City University of London, Department of Mathematics, London EC1V 0HB, UK}
\affil[2]{\small The Alan Turing Institute, British Library, 96 Euston Road, London NW12DB, UK}
\affil[3]{\small UCL Centre for Blockchain Technologies, University College London, London, UK.}

\affil[*]{\small abaronchelli@turing.ac.uk}

\begin{document}

\maketitle

\textbf{Blockchains are still perceived chiefly as a new technology. 
But each blockchain is also a community and a social experiment, built around social consensus. Here I discuss three examples showing how collective intelligence can help, threat or capitalize on blockchain-based ecosystems. They concern the immutability of smart contracts, code transparency and new forms of property. The examples show that more research, new norms and, eventually, laws are needed to manage the interaction between collective behaviour and the blockchain technology. Insights from researchers in collective intelligence can help society rise up to the challenge. }

For most people, blockchain technology is about on-chain consensus. And it certainly is. A blockchain provides users with a method and incentives for validating and storing a value or transaction without the need to trust a central authority, or any other participants in the system. Once approved, the transaction is recorded in an open ledger that cannot be altered retroactively. Since the system is trustless, everyone can join the network and read, write, or participate within the blockchain (in this article, I am considering only public and permissionless blockchains) \cite{kube2018daniel}.

But a public blockchain is not a self-contained universe. As a socio-technical system, its life extents behind the user base. For example, a blockchain can provide competition to existent socio-economic players, and it is shaped by what users do with it and, eventually, by the law \cite{baronchelli2018emergence,de2018blockchain}. Continuous updates and strategic decisions are therefore required to keep up with such a rapidly evolving landscape, meaning that blockchains are not self-sufficient. This has two implications. The first is that blockchains need governance, defining how the community of users, developers and immediate stakeholders can (agree on how to) make decisions \cite{reijers2016governance,govbase}. The second implication is that, in order for those decisions to be meaningful, the governing community needs to have a clear understanding of the social experiment they have launched, i,e., how norms and behaviours are impacting, and will be impacted by, the blockchain itself. 

Considering that each blockchain is also a community and a social experiment is key to understand the evolution and impact of the technology~\cite{new_yorker}. Yet it is often overlooked both by blockchain stakeholders, who may be too absorbed by the technical aspects, and by researchers, who have traditionally been slow to detect societal transformation. The latter is unfortunate since most data are public and the challenges raised by the blockchain technology, even just in the form of cryptocurrency, have been increasing steadily. Here, I focus on three examples showing what can happen when collective intelligence meets the blockchain, and the design and incentives built in the latter turn out to be incomplete or problematic. The first two examples concern blockchains and their communities, the last one is about the unpredicted fortune of what is now a large-scale socio-economic revolution. In all cases, I will highlight the challenges for researchers and society as a whole.

\subsection*{Coping with the (im)mutability of smart contracts} Who makes decisions \textit{about} a blockchain? This question, initially overlooked, became urgent as soon as Bitcoin gained the first users. An efficient governance guarantees that a blockchain keeps offering its users a competitive environment \cite{dupont2017experiments,beck2018governance,de2018blockchain,govbase}. Most companies and organisation adopt a centralised governance system with a leadership team at its top. However, blockchain systems tend to prefer decentralisation almost by nature. Thus, governance becomes a community problem, and every decision - including the decision on how to make decisions - a consensus problem in itself.

The complex interplay between governance and decentralisation manifested clearly in June 2016, when a large amount of ethers suddenly “disappeared” from the Ethereum ecosystem \cite{dupont2017experiments,dao_wiki}. They got hacked from a complex smart contract known as “The DAO”. Described “as a new paradigm of economic cooperation [..] a digital democratization of business” \cite{daotechcrunch}, i.e. a grandiose social experiment, the DAO had been funded with \$120 millions obtained through a crowdfunding campaign \cite{daocnbc}. The DAO contained around 15\% of all ethers, and the hack affected approximately 30\% of those, i.e. 5\% of all available ethers.

The size of the hack put the whole Ethereum ecosystem under major stress, undermining its credibility. The price of ether fell by more than $35\%$ \cite{daocoindesk_sol}. Furthermore, Ethereum aimed (and still aims) at switching to a proof of stake governance model, where the mining or validating power is proportional to the owned amount of coins, so an entity with 5\% of all ethers would become extremely influential.

The community had several possibilities \cite{daocoindesk_sol} including (1) Accepting the hack and its consequences, including the possible devaluation of the ecosystem. This would potentially imply the end of Ethereum and the loss of a lot of real money; (2) Blocking the account of the hacker. This proposal, dubbed “soft fork”, was extensively discussed by the community, but had several problems, including legal risks; or (3) Rewriting history by invalidating all the transactions that followed the hack and start a parallel universe where the hack had never occurred and investors would get their money back (blockchains, parallel universes and rewriting history is a fascinating topic but unfortunately outside of our scope here).

Option (3) would make it possible for everyone who participated in the DAO to withdraw their funds. With the support of the miners, and because nothing had been spent so far, nothing would be lost. Critics warned that rewriting history would undermine the perception that blockchains are immutable, with devastating consequences for Ethereum. Yet the community chose to proceed this way, performing an unprecedented “hard fork”. Ethereum went back to square one, to before the hack. A minority of “rebels”, approximately 15\% of those who were involved in the decision process, continued instead using the unaltered original blockchain, now called “Ethereum Classic”, which has survived to today \cite{eth_classic}. The operation was successful. Ethereum is still in good health, ranking second for market capitalisation behind Bitcoin. There was no drop in price after the hack and, in fact, a few months later Ethereum became one of the main characters of the 2017 boom of cryptos. 

The DAO story shows the importance of off-chain decisions on the life of a blockchain. Consensus, or lack thereof, among developers, governance, users and miners does affect the economic interests of thousands  of individuals. It also highlights how society and researchers need to reflect on how to deal with the immutability of smart contracts (not only in relation to the foreseeable and partially already started proliferation of DAOs \cite{beck2018governance}), in order to confront bugs or changed circumstances that may make the contract obsolete. For ``code is law''~\cite{lessig1999code} to be an opportunity and not a prison, we need new norms. This requires both more research and a broader societal discussion.

\subsection*{The role of developers and code transparency as a systemic property}
The classic risk associated to the ``code is law'' approach~\cite{lessig1999code} is that developers would acquire too much power~\cite{natl_law_rev}. The accepted antidote has been to open up the code. Open source code would limit the influence of developers by allowing everyone to monitor what they write~\cite{natl_law_rev}. Of course, the assumption that someone will in fact (be able to) read the code can appear a bit naive, but the mere possibility to read it would be enough to guarantee transparency. This argument is crucial to crytpocurrencies, at present the most advanced use case of blockchains. Every cryptocurrency is entirely defined by its code, which determines its security, functionality, availability, transferability, and general malleability~\cite{antonopoulos2014mastering}. Hence, the code of most of them is open source. 

However, the open source argument has a weakness. It assumes that cryptocurrencies (or, in general, any coded objects) are isolated entities, each one living in its own silo. Starting from this observation, my collaborators and I looked at cryptocurrency codes stored in GitHub and found that 4\% of developers contributed to the code of two or more cryptocurrencies \cite{lucchini2020code}. This group included some of the most active coders. Taken together, they are responsible for 10\% of all edits. Hence, at the code level, cryptocurrencies are not independent. Does this matter?

The cryptocurrency market is the natural place to assess the impact of developers working on more than one project. Since cryptos are traded, their price trends can be used to compare their life beyond the code. It turns out that the temporal evolution of the network of co-coded crypto-currencies anticipates market behaviour. In particular, the first time two independent codes get connected via the activity of one shared developer marks, on average, a period of increased correlation between the returns of the corresponding cryptocurrencies  \cite{lucchini2020code}.

Of course, such an analysis can not reveal `why' such correlation is established. We can speculate, for example, that developers are either directly `rich' in one of the two cryptocurrencies or employed by some wealthy actor, and that their activity correlates with some financial interest (typically pairs of co-developed cryptos include a major and a smaller coin). However, the point is that the correlation between returns of co-coded cryptos would be virtually impossible to discover if we ignored the collaborative activity of the developers, and that correlations are important to consider for example when assessing the risks of an investment portfolio. 

In this example, the main actors were developers but it is easy to see how governance could have an even more dramatic impact. Cryptos are entirely defined by their code, true. But their behaviour - in this case their market price - is shaped also by the community that builds them. Future research clarifying how this may happen will have an impact that transcends cryptocurrencies. Understanding whether and how financial markets and technological - code - development interact is an open and debated question~\cite{4_github_software_repository,5_innovation_potential_in_cryptocurrencies,6_bitcoin_effects_in_return,7_cryptocurrencies_as_financial_asset}. More broadly,  today several code-based ecosystems challenge traditional institutions, from national laws to financial markets~\cite{lessig2007code,de2018blockchain,16_cryptocurrencies_and_blockchain} and code transparency will be more and more crucial to prevent societal risks.

\subsection*{How a blockchain game revolutionised the art market}

Fungibility is a crucial feature of money, where individual units must be interchangeable. One euro coin is identical, from the practical point of view, to any other euro coin. The same is true for one gram of gold. Since the blockchain was originally conceived to enable electronic cash, fungibility was not questioned. At least until 2014, when the first experiments with non-fungibility were proposed to the community of Ethereum developers \cite{NFT_end}; and then 2017, when Non Fungible Tokens (NFTs) gained popularity in the form of a an Ethereum game called Cryptokitties~\cite{kitties_main_page}, that  allowed  players  to  purchase,  collect,  breed  and  sell virtual  cats. 

The game was broadly considered a symptom of the `cryptocurrency madness'~\cite{kitty_madness}, but the underlying technology had some interesting properties. At its roots, an NFT is a smart contract that points to some other data, like an image or video. Living on a public blockchain, predominantly Ethereum, NFTs allow everyone to verify their authenticity and their past history. After years of relative calm, the NFT attracted major media attention in March 2021, when the artist Beeple sold an NFT of his work for \$69.3 million at Christie's~\cite{christie_nft_sold} (the third-highest auction price achieved for a living artist, after Jeff Koons and David Hockney~\cite{beeple_third}). Several other multi-million sales followed~\cite{Most_expensive_sales}. NFTs profitability has attracted several celebrities, who created their own NFTs, as well as the most popular sports, with collectibles of NBA and famous football players that are currently sold for hundreds of thousands dollars~\cite{Trading_collectibles_NFT}. At the moment of writing, the market has surpassed \$1 billion \cite{NFT_noi} and it will likely grow even further as NFTs are being used to commodify digital objects in different contexts beyond art, such as gaming (the first industry to make profit of the technology), sport collectibles, music and even physical objects such as fashion items \cite{NFT_noi}.

Collective intelligence expanded the original idea associated to the blockchain technology by dropping fungibility, a key feature of Bitcoin. The new NFT technology was then perfected through a series of protocol improvements \cite{} and eventually landed in - and revolutionised - unforeseen territories, starting with digital art. As NFTs are now conquering also physical objects, new norms and regulations will have to decide how they interact with existent property and trade laws. The social experiment has gone well. It is time for society to make the most of it.

\subsection*{Conclusion}  The above examples highlight some of the challenges and opportunities raised by the blockchain technology, namely the immutability of code-based contracts, the concentration of power in the hands of few code-savvy individuals, and the friction between new social norms and existing laws to define property. It is easy to see that more challenges will come in the future, and many others are already with us. The understanding of the interplay between collective behaviour and the technology will help design better incentives and better blockchains, but it is difficult to imagine that design will ever be able to anticipate everything. For now, we need grounded answers for questions such as when decentralisation offers real advantages, how we can help design new social norms, how law and norms interact, what the impact of code on society is, how unregulated markets emerge and self-organise, how several currencies can coexist, and what level of individual identity or anonymity is needed to guarantee social coordination and everyday transactions. Many of these issues - in different forms - have long been familiar to researchers in collective intelligence, and the field can give a tremendous contribution towards understanding and shaping the imminent blockchain revolution.

\end{document}